\documentclass[pre,floatfix,superscriptaddress,preprint,showpacs]{revtex4}
\usepackage{latexsym}
\usepackage{amstext,amsfonts,amssymb,amsmath}
\usepackage{tabularx}
\def\sigmaconf{\{\boldsymbol{\sigma}\}_{\rm conf}}

\def\sigmabold{\boldsymbol{\sigma}}
\def\pbold{\boldsymbol{\rm p}}
\def\Oboldij{\boldsymbol{\rm O}_{ij}}
\def\TboldNr{\boldsymbol{\rm T}(N,r)}
\def\TboldNrij{\boldsymbol{\rm T}_{ij}(N,r)}
\def\TbarboldNrij{\overline{\boldsymbol{\rm T}}_{ij}(N,r)}
\def\TboldNri{\boldsymbol{\rm T}_i(N,r)}

\usepackage{epsfig}
\begin{document}
\title{Weighted projected networks: mapping hypergraphs to networks}
\author{Eduardo L\'{o}pez}
\affiliation{CABDyN Complexity Centre, Sa\"{\i}d Business School, University of Oxford, Park End Street,
Oxford OX1 1HP, United Kingdom}
\affiliation{Physics Department, Clarendon Laboratory, University of Oxford, Parks Road, Oxford OX1 3PU,
United Kingdom}
\email{eduardo.lopez@sbs.ox.ac.uk}
\date{\today}
\begin{abstract}
Many natural, technological, and social systems incorporate multiway interactions, 
yet are characterized and measured on the basis of weighted pairwise interactions. In this article, I 
propose a family of models in which pairwise interactions originate from multiway interactions, by
starting from ensembles of hypergraphs and applying projections  
that generate ensembles of {\it weighted projected networks}.
I calculate analytically the statistical properties of weighted projected networks, and suggest ways
these could be used beyond theoretical studies. Weighted projected networks typically exhibit weight disorder 
along links even for very simple generating hypergraph ensembles.
Also, as the size of a hypergraph changes,
a signature of multiway interaction emerges on the link weights of weighted projected networks that distinguishes them 
from fundamentally weighted pairwise networks.
This signature could be used to search for hidden multiway interactions in weighted network data.
I find the percolation threshold and size of the largest component for hypergraphs
of arbitrary uniform rank, translate the results into projected networks, and show that the transition
is second order.
This general approach to network formation has the potential to shed new light on our understanding of
weighted networks.
\end{abstract}
\maketitle
\section{Introduction}
Recent years have seen the growth of complex networks theory, a research area 
concerned with the general theory of systems of interacting elements~\cite{rev-Albert}. 
Its relevance has been illustrated in a number of problems, such as 
infectious disease propagation~\cite{Colizza}, 
the strength of social ties~\cite{Onnela}, 
data routing in technological networks~\cite{Sreenivasan}, 
and motifs in biological networks~\cite{Shen-Orr}. 
An underlying driver for the growth of this field has been the increased availability of digitized
information, which can be efficiently analyzed to uncover relations between system elements.

A simplifying assumption that is made in networks theory is to characterize interactions as
being exclusively pairwise (each interaction represented by a link between two nodes), often with 
an associated interaction intensity or weight, generating so-called weighted networks
(also known as weighted graphs in Mathematics). 
The reason for this approach is that usually the information available for real systems is relatively limited. 
Despite these limitations, weighted networks have proven very useful, as a number of
measurable network quantities have shown their relevance in application.
Examples of these quantities are the distribution of node degree~\cite{Albert-perc,Cohen}
(number of links connecting to a node),
optimal path lengths between network nodes~\cite{Chen}, and node clustering~\cite{Watts}
(a measure of loops of length three).
Other properties that depend on specific groups of links (e.g., network communities)
have also proven quite useful~\cite{Fortunato,Porter}.

There are situations, however, where it is known that interactions extend to groups larger than two
(multiway interactions), and one can use such information to create more accurate models, 
avoiding the possibility of oversimplified or misleading results. 
Examples of these situations are, for instance, networks of affiliations~\cite{Wasserman,Borgatti,Wang},
where nodes representing individuals connect to each other by virtue of their membership to a group
such as their family or workplace colleagues; another example
are folksonomies~\cite{Ghoshal}, systems that encode information of triplets of the following
three ingredients: objects, descriptors of the objects, and the individuals making the descriptions.
Characterizing these examples by avoiding the pairwise simplification should lead to more
informative and reliable results. 

Through various independent approaches, researchers focusing on problems of multiway 
interactions have proposed mechanisms by which pairwise network
weights are generated as a consequence of these interactions (see, e.g.~\cite{Wasserman} and \cite{fn-bipartite}). 
For instance, in affiliation
networks, when two nodes belong simultaneously to multiple groups, a feature called co-membership,
it is assumed that their relationship intensity is equal to the number of groups they both belong to.
Perhaps the most appealing feature of these ideas is that they provide a unifying principle to
the structure of some interacting systems: the presence of a group generates links, and being part
of multiple groups generates weights.
Surprisingly, these unifying ideas have received limited attention, perhaps because
some of the mathematical models that are required are less straightforward than typical networks.
Here, I focus on a systematic approach grounded in statistical mechanics to relate multiway interactions 
to weighted networks.

To model multiway interactions, it is appropriate to use hypergraphs, which are generalizations
of networks~\cite{hypergraphs}. They are composed of a set of nodes and a set of hyperedges. Each hyperedge 
is a group of interconnected nodes (a clique), and the hypergraph is the collection of all the hyperedges
and isolated nodes; networks are the specialization of hypergraphs in which all hyperedges are
cliques each with only two nodes, i.e., links. The size of a hyperedge is called rank. 
In a statistical mechanics formulation (random) hypergraphs are called homogeneous when all hyperedges are equally likely
to be present, or heterogeneous when each hyperedge has its own (possibly unique) probability to appear. 
For the examples mentioned above: in a folksonomy, for instance, hyperedges are all
of rank three, whereas in affiliation networks, in principle, hyperedges can have different ranks; 
both examples are likely to be heterogeneous hypergraphs. 

The notion of hypergraphs generating weights is equivalent to constructing networks that represent
a projection of a hypergraph. In other words, starting from a hypergraph, one can create an associated
set of links that form a {\it weighted projected network}, where each link weight is given by the 
structure of the hypergraph and a projection rule. This construction suggests some intriguing possibilities:
some data that is typically studied as a network may in fact emerge from underlying hypergraphs.
If that is the case, it should be possible in principle to construct hypergraph models and accompanying 
projections that can fit observed data and narrow down its origins.

In this article, I study homogeneous and heterogeneous entropy maximizing hypergraph ensembles 
of arbitrary uniform rank $r$ and define general projections of hypergraphs that lead to ensembles 
of weighted projected networks.
Some specific
projection examples that have been used in the literature are explored~\cite{Wasserman,fn-bipartite}, the properties
of their respective projected networks calculated, and their interpretations briefly discussed.
The percolation threshold and size of the largest connected component of hypergraphs of arbitrary uniform rank 
are also derived by use of the mapping between the Potts model and percolation theory~\cite{Fortuin}, 
and the results are then translated into the percolation properties of the projected networks. These results
show that the transition is of second order.
I find that, as a function of size, the link weights on weighted projected networks
can display a signature of the presence of hidden multiway relations:
when faced with a weighted network, this signature could provide indications that
there is an associated hypergraph hidden in the data.

The article is structured in the following way: Sec.~\ref{max-ent} focuses on the general definitions 
of projections of hypergraphs onto networks, and on models of entropy maximizing ensembles of hypergraphs.
With these results, in Sec.~\ref{applications} I study in greater detail the statistical properties of 
general projected networks, as well as some concrete examples.
These results suggest how to explore network data for possible signatures of multiway relations.
Completing the results, Sec.~\ref{percolation} focuses on the percolation properties 
of hypergraphs and their projected networks, and explores the general notion of sparsity. 
I finalize the article in Sec.~\ref{conclusions} with some discussion and conclusions.
\section{Maximum entropy hypergraphs and the network projection}
\label{max-ent}
Consider a hypergraph, represented by $\sigmabold$, consisting of a set of nodes $1,\dots,N$,
and for each possible hyperedge of $r$ nodes $i_1,\dots,i_r$, an indicator 
$\sigma_{i_1,\dots,i_r}$ equal to 1 if the hyperedge is present and 0 if it is absent; all 
subindices $i_1,\dots,i_r$ take non-repeated values from the set $\{1,\dots,N\}$. 
In general, a hypergraph does not require $r$ to be the
same for all hyperedges. However, for the sake of simplicity, I focus on single rank 
(all hyperedges have the same $r$) undirected hypergraphs, with the indicator
$\sigma_{i_1,\dots,i_r}$ symmetric under permutations of $i_1,\dots,i_r$ (if one is interested in
studying combinations of rank, one merely requires the introduction of the proper parameters for this,
but the qualitative nature of the problem is the same as that studied here). 
Unweighted undirected networks correspond to $r=2$.

The general hypergraph projection onto a network is defined as a function $\mathcal{P}$ applied
over hyperedges of $\sigmabold$ that produces the adjacency matrix $w_{ij}$ 
for the weighted projected network $G$. Network $G$ is formed by the same node set as $\sigmabold$, and its 
adjacency matrix is $w_{ij}$. If a node does not belong to any hyperedge, it is isolated in both
$\sigmabold$ and $G$.
For given $\sigmabold$, one can define the subset 
$O_{ij}(\sigmabold):= \{(i_1,\dots,i_r)|(i_1,\dots,i_r)\in\sigmabold
\wedge i\in \{i_1,\dots,i_r\} \wedge j\in \{i_1,\dots,i_r\}\}$ of its hyperedges
that include simultaneously nodes $i$ and $j$. 
The kinds of projections studied here are of the type
\begin{equation}
w_{ij}(G)=\mathcal{P}\left(\left|O_{ij}(\sigmabold)\right|\right),
\label{wij-P}
\end{equation}
where $o_{ij}\equiv\left|O_{ij}(\sigmabold)\right|$ is the size (cardinality) of $O_{ij}(\sigmabold)$. 
Thus, the weight of link $ij$ in $G$ only depends on the number of hyperedges that contain $i$ and $j$,
an intuitive choice, although certainly not the only possible model (in the literature, all examples
I have found are limited to this kind of projection~\cite{Wasserman,fn-bipartite,Yoon}).

\begin{figure}
\epsfig{file=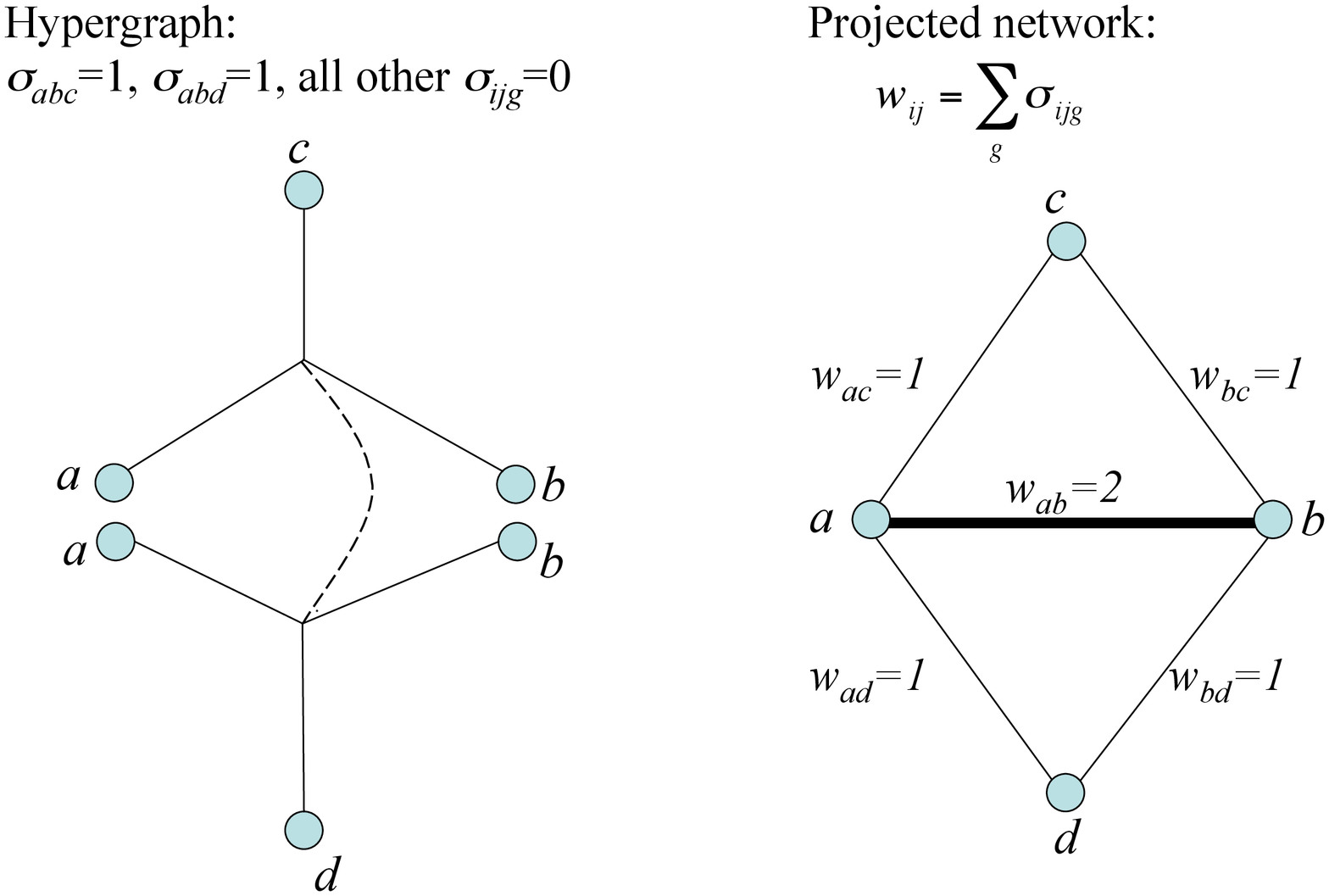,angle=270,scale=0.5}
\caption{Illustration for the projection $\mathcal{P}_a$ from hypergraphs to networks. On the left, a hypergraph
is composed of a multitude of hyperedges that exist when $\sigma=1$, and do not when $\sigma=0$.
The projected network (right) has a link between all nodes that belong to the same hyperedge, and 
the weight of the link is the number of hyperedges that share the same pair of nodes.}
\label{wij-project-illustration}
\end{figure}

On a concrete empirical case, the projection $\mathcal{P}$ should reflect the understanding of the
relation between $\sigmabold$ and $G$.
Here, I present results for some reasonable sample choices 
of $\mathcal{P}(O_{ij})$, namely $\mathcal{P}_a(O_{ij})=o_{ij}$ (additive projection) 
and $\mathcal{P}_n(O_{ij})=\theta(o_{ij})$ (nominal projection),
where $\theta$ is the Heaviside step function ($=0$ if the argument is 0 or less, and 1 otherwise).
In addition, I show some features satisfied by the projected networks generated by a large class of
projections with the general form of Eq.~(\ref{wij-P}).
To perform calculations, note that
the additive projection can be written in terms of $\sigma_{i_1,\dots,i_r}$ as
\begin{equation}
\mathcal{P}_a(O_{ij}(\sigmabold))=o_{ij}=\sum_{(i_1,\dots,i_r)\in O_{ij}(\sigmabold)} \sigma_{i_1,\dots,i_r},
\end{equation}
whereas the nominal projection is represented by
\begin{equation}
\mathcal{P}_n(O_{ij}(\sigmabold))=\theta\left(\sum_{(i_1,\dots,i_r)\in O_{ij}(\sigmabold)} 
\sigma_{i_1,\dots,i_r}\right).
\end{equation}
An illustration of $\mathcal{P}_a$ for the case of $r=3$ is shown in Fig.~\ref{wij-project-illustration}.

In the literature, both hypergraphs and projections have been used to study interaction data
qualitatively embedded in complex networks theory, but without a sense of unification.
For instance, the choice $\mathcal{P}_n(O_{ij})$
is implicit in work such as~\cite{Yoon}; there, if $\sigma_{i_1,\dots,i_r}$ is interpreted as a specific motif
(structural pattern), the model generates unweighted networks guaranteed to posses those motifs.
In another approach, found in Refs.~\cite{Newman-clusters,Miller-clusters}, 
each hypergraph (containing $r=2$ and 3 only) treats each rank separately in that 
the interactions of nodes by way of pairs is counted independently to the triplet interactions,
with no notion of projection onto a simple network.
Refs. \cite{Ghoshal, Wasserman} do consider projections in some form, but are limited
by rank $r$ of hypergraph and by the nature of the projection.
Projection $\mathcal{P}_a$ is in fact common~\cite{Wasserman}, and it is often used as
a way to characterize the one-mode networks that emerge from bipartite graphs~\cite{fn-bipartite} 
(recent work also uses the notion of projection in the context of time evolving networks~\cite{Barrat}).
Eq.~(\ref{wij-P}) offers a unified way to relate networks and hypergraphs, which can be
applied to the models cited above to develop additional understanding of the problems.

To build unbiased statistical models, I adapt to hypergraphs the canonical ensemble
approach developed in Ref.~\cite{Park}.
The set of all possible hypergraphs $\sigmabold$ is given by
$\sigmaconf$ (the ensemble), or in other words, $\sigmaconf$ is the union
of all possible unique hypergraphs $\sigmabold$.
To analytically formulate the ensemble problem, consider the entropy $S$, defined as
\begin{equation}
S=-\sum_{\sigmaconf} P(\sigmabold)\ln P(\sigmabold),
\end{equation}
where $P(\sigmabold)$ represents the probability
of a given configuration within the hypergraph ensemble, and the sum
over configurations is equivalent to summing over all hyperedge combinations, or
$\sum_{\sigmaconf}\rightarrow \sum_{\sigma_{1,\dots,r}=0}^1\dots\sum_{\sigma_{N-r+1,\dots,N}=0}^1$.
The canonical ensemble
approach finds the distribution $P(\sigmabold)$ that maximizes $S$ while satisfying conditions that define
the ensemble of interest. Such conditions, say $\{\langle X_{\alpha}\rangle\}$, with
$\alpha$ an enumeration index, are taken to be of the form
\begin{equation}
\sum_{\sigmaconf}X_{\alpha}(\sigmabold)P(\sigmabold)=
\langle X_{\alpha}\rangle.
\end{equation}
Finally, since $P(\sigmabold)$ are probabilities, one must guarantee 
normalization, which translates into
\begin{equation}
\sum_{\sigmaconf}P(\sigmabold)=1.
\end{equation}
The solution to this problem ($P(\sigmabold)$ satisfying the conditions above) 
is obtained via Lagrange multipliers. Each condition is 
related to a multiplier, and one solves the equations 
\begin{equation}
\frac{\partial}{\partial P(\sigmabold)}\left[
S+\eta\left(1-\sum_{\sigmaconf}P(\sigmabold)\right)+
\sum_{\alpha}\beta_{\alpha}\left(\langle X_{\alpha}\rangle-\sum_{\sigmaconf}
X_{\alpha}(\sigmabold)P(\sigmabold)\right)
\right]=0,
\label{lagrange-problem}
\end{equation}
for $P(\sigmabold)$, with $\eta, \beta_1,\dots$ the Lagrange multipliers. 
The solution to the problem can be expressed as
\begin{equation}
P(\sigmabold)=\frac{e^{-H(\sigmabold)}}{Z}.
\label{lagrange-solution}
\end{equation}
The partition function $Z$, and $H(\sigmabold)$ ({\it defined} as the Hamiltonian),
are respectively given by
\begin{equation}
Z=\sum_{\sigmaconf}e^{-H(\sigmabold)}
\end{equation}
and 
\begin{equation}
H(\sigmabold)=\sum_{\alpha}\beta_{\alpha}X_{\alpha}(\sigmabold).
\label{hamiltonian}
\end{equation}

Among the simplest non-trivial problems one can address is that of the fully random
hypergraph with equal probability for any hyperedge to exist. The constraint associated
with this example is the requirement that there is a given average number of 
hyperedges, $\langle L_r\rangle$, over the hypergraph ensemble. 
Since $L_r$ for a given configuration $\sigmabold$
is given by $L_r(\sigmabold)=\sum_{(i_1,\dots,i_r)\in \sigmabold} \sigma_{i_1,\dots,i_r}$,
the set of constraints reduces to two Lagrange multipliers, one for
the normalization, and another parameter, labelled $\beta$, for $\langle L_r\rangle$. 
Introducing this in Eq.~(\ref{lagrange-problem})
generates the Hamiltonian
\begin{equation}
H(\sigmabold)=\beta\sum_{(i_1,\dots,i_r)\in\sigmabold}
\sigma_{i_1,\dots,i_r}=\beta L_r(\sigmabold)
\end{equation}
and the partition function
\begin{multline}
Z=\sum_{\sigmaconf}e^{-\beta\sum_{(i_1,\dots,i_r)\in\sigmabold}\sigma_{i_1,\dots,i_r}}\\
= \sum_{\sigma_{1,\dots,r}=0}^1\dots\sum_{\sigma_{N-r+1,\dots,N}=0}^1 
\prod_{(i_1,\dots,i_r)\in\TboldNr}e^{-\beta\sigma_{i_1,\dots,i_r}}\\
=\sum_{\sigma_{1,\dots,r}=0}^1 e^{-\beta\sigma_{1,\dots,r}}
\dots\sum_{\sigma_{N-r+1,\dots,N}=0}^1 e^{-\beta\sigma_{N-r+1,\dots,N}}
=(1+e^{-\beta})^{{N\choose r}},
\end{multline}
where $\TboldNr$ is the set of all possible hyperedges $\{(1,\dots,r),\dots,(N-r+1,\dots,N)\}$,
i.e., the complete hypergraph of single rank $r$ and size $N$.
The last equality can also be obtained from the symmetry of the Hamiltonian over exchange 
of indices among $\sigma_{i_1,\dots,i_r}$.
The result expresses that there are ${N\choose r}$ possible hyperedges $(i_1,\dots,i_r)$ 
among the $N$ nodes. 
Using this result one can show that the $\langle L_r\rangle$ constraint is satisfied for the
proper choice of $\beta$, as seen from
averaging $L_r(\sigmabold)$ in the $P(\sigmabold)$ ensemble
\begin{equation}
\langle L_r\rangle=\frac{1}{Z}\sum_{\{\boldsymbol{\sigma}\}_{\rm conf}}
\sum_{(i_1,\dots,i_r)\in\sigmabold}\sigma_{i_1,\dots,i_r} 
e^{-\beta\sum_{(j_1,\dots,j_r)\in\sigmabold}\sigma_{j_1,\dots,j_r}}
=\frac{1}{Z}\sum_{L_r=0}^{{N\choose r}}{{N\choose r}\choose L_r}L_r e^{-\beta L_r}={N\choose r}p,
\end{equation}
and $p\equiv (1+e^\beta)^{-1}$ is the probability for a hyperedge to be present, which is evident
from writing $\langle L_r\rangle/{N\choose r}=(1+e^\beta)^{-1}=p$; $p$ also corresponds to the expectation
value of any hyperedge 
$\langle \sigma_{i_1,\dots,i_r}\rangle=\sum_{\sigmaconf}\sigma_{i_1,\dots,i_r}P(\sigmabold)=
(1+e^\beta)^{-1}=p$, i.e., the probability for any hyperedge to exist.
The fact that all hyperedges are equally likely suggests referring to this case as the {\it homogeneous
hypergraph ensemble}.
The probability of a specific hypergraph configuration to be observed is given by 
Eq.~(\ref{lagrange-solution}), which in this case yields
\begin{equation}
P(\sigmabold,p)=p^{L_r(\sigmabold)}(1-p)^{{N\choose r}-L_r(\sigmabold)}
\end{equation}
where the relations $1+e^{-\beta}=(1-p)^{-1}$ and $e^{-\beta}=p(1-p)^{-1}$ have been used.
The application of the $\mathcal{P}_a$ and $\mathcal{P}_n$ to the homogeneous ensemble is tackled below
in a more general ensemble.

The solution to the simple homogeneous problem above, helps to identify some basic features
of the canonical approach, including quantities such as the probability of a hyperedge, and of a
specific hypergraph $\sigmabold$. Building on this, one can construct the more general {\it heterogeneous}
case, where each hyperedge has its own expectation $\langle\sigma_{i_1,\dots,i_r}\rangle$. Thus, the
Hamiltonian of Eq.~(\ref{hamiltonian}) becomes
\begin{equation}
H(\sigmabold)=\sum_{(i_1,\dots,i_r)\in\sigmabold}\beta_{i_1,\dots,i_r}\sigma_{i_1,\dots,i_r}.
\end{equation}
In analogy with the homogeneous case, one defines $p_{i_1,\dots,i_r}\equiv\langle\sigma_{i_1,\dots,i_r}
\rangle =(1+e^{\beta_{i_1,\dots,i_r}})^{-1}$.
The partition function becomes
\begin{equation}
Z(\pbold)=\prod_{(i_1,\dots,i_r)\in\TboldNr}(1+e^{-\beta_{i_1,\dots,i_r}})
=\prod_{(i_1,\dots,i_r)\in\TboldNr}(1-p_{i_1,\dots,i_r})^{-1},
\end{equation}
where $\pbold$ represents the hyperedge expectations $\{p_{1,\dots,r},\dots,p_{N-r+1,\dots,N}\}$.
The probability of a hypergraph configuration $\sigmabold$ is then
\begin{equation}
P(\sigmabold,\pbold)=\prod_{(i_1,\dots,i_r)\in\TboldNr}p^{\sigma_{i_1,\dots,i_r}}_{i_1,\dots,i_r}
(1-p_{i_1,\dots,i_r})^{1-\sigma_{i_1,\dots,i_r}},
\label{Psigma_hetero}
\end{equation}
which is the joint probability that hyperedges with $\sigma_{i_1,\dots,i_r}=1$ are present,
and those with $\sigma_{i_1,\dots,i_r}=0$ are absent. 
If for all $(i_1,\dots,i_r)\in \TboldNr$, $p_{i_1,\dots,i_r}=p$, one recovers the homogeneous case.
The heterogeneous ensemble possesses the most degrees of freedom among non-interacting undirected hypergraph
models. If more specific constraints are imposed such as, for instance, conditions on the
average number of hyperedges visiting a node, they would translate into additional constraints on
the values of the set $\pbold$ (see e.g. discussion at end of Sec.~\ref{applications}).
\section{Application of the hypergraph projection}
\label{applications}
Since only projections of the form $\mathcal{P}(O_{ij})=\mathcal{P}(o_{ij})$ are considered here,
the statistical properties of the projected networks depend on the statistical properties of $o_{ij}$. 
It is most useful to focus on the distribution of $o_{ij}$ in the heterogeneous ensemble, 
$\phi_{ij}(o_{ij},\pbold)$, and determine how this translates into the homogeneous case
(Table~\ref{table-phi_oij} summarizes the notation used to compute $\phi_{ij}(o_{ij},\pbold)$).
Let us define $\TboldNrij$ as the set of all hyperedges on the complete hypergraph that visit
$i$ and $j$ simultaneously.
I also define $\TbarboldNrij$, the complement of $\TboldNrij$ with respect to $\TboldNr$.
In  addition, for each configuration $\sigmabold$, $V_{ij}(\sigmabold)$ is the set of hyperedges visiting 
$i$ and $j$ which may or may not have cardinality $o_{ij}$ (if it does, it is represented as before
with $O_{ij}(\sigmabold)$). The complement of $V_{ij}(\sigmabold)$ with respect to $\TboldNrij$
is $\overline{V}_{ij}(\sigmabold)$, and thus $\TboldNrij=V_{ij}(\sigmabold)\bigcup\overline{V}_{ij}(\sigmabold)$.
Taking this into account, $\phi_{ij}(o_{ij},\pbold)$ can be calculated through the expression
\begin{equation}
\phi_{ij}(o_{ij},\pbold)=\sum_{\sigmaconf}
\delta\left(o_{ij},\sum_{(i_1,\dots,i_r)\in V_{ij}(\sigmabold)} 
\sigma_{i_1,\dots,i_r}\right)P(\sigmabold,\pbold),
\label{phi_oij_def}
\end{equation}
where $\delta(x,y,\dots,z)$ is the Kronecker delta which can have two or more arguments, 
and is equal to 1 if all the arguments are equal, and 0 otherwise.
In the sum above, only those configurations for which delta is 1 contribute to $\phi_{ij}(o_{ij},\pbold)$,
and this occurs only when there are exactly $o_{ij}$ hyperedges in $\sigmabold$ 
that include $ij$. 

To perform the calculation, note the independence of each component of $\pbold$ in
Eq.~(\ref{Psigma_hetero}). This allows factoring the sum over configurations in Eq.~(\ref{phi_oij_def})
into a product of i) the configurations of hyperedges $\TbarboldNrij$,
which cannot affect the delta, and ii) the configurations of hyperedges $\TboldNrij$, which can.
The hyperedges 
$(i_1,\dots,i_r)\in\TbarboldNrij$
each contribute a factor 
$\sum_{\sigma_{i_1\dots,i_r}=0}^1 p_{i_1,\dots,i_r}^{\sigma_{i_1,\dots,i_r}}
(1-p_{i_1,\dots,i_r})^{1-\sigma_{i_1,\dots,i_r}}=1$.
Therefore, the remaining factors of Eq.~(\ref{phi_oij_def}) lead to
\begin{multline}
\phi_{ij}(o_{ij},\pbold)=\sum_{V_{ij}(\sigmabold)\subset \boldsymbol{\rm V}_{ij}}
\delta\left(o_{ij},\sum_{(i_1,\dots,i_r)\in V_{ij}(\sigmabold)} 
\sigma_{i_1,\dots,i_r}\right)
\prod_{(i_1,\dots,i_r)\in \TboldNrij}
p_{i_1,\dots,i_r}^{\sigma_{i_1,\dots,i_r}}(1-p_{i_1,\dots,i_r})^{1-\sigma_{i_1,\dots,i_r}}\\
=\sum_{V_{ij}(\sigmabold)\subset \boldsymbol{\rm V}_{ij}}
\delta\left(o_{ij},\sum_{(i_1,\dots,i_r)\in V_{ij}(\sigmabold)} 
\sigma_{i_1,\dots,i_r}\right)
\prod_{(i_1,\dots,i_r)\in V_{ij}(\sigmabold)} p_{i_1,\dots,i_r}
\prod_{(i_1,\dots,i_r)\in \overline{V}_{ij}(\sigmabold)} (1-p_{i_1,\dots,i_r})\\
=\sum_{O_{ij}(\sigmabold)\subset \Oboldij}
\prod_{(i_1,\dots,i_r)\in O_{ij}(\sigmabold)} p_{i_1,\dots,i_r}
\prod_{(i_1,\dots,i_r)\in \overline{O}_{ij}(\sigmabold)} (1-p_{i_1,\dots,i_r}),
\label{phi_oij_final}
\end{multline}
where $\boldsymbol{\rm V}_{ij}$ and $\Oboldij$ are the unions of all possible sets $V_{ij}(\sigmabold)$
and $O_{ij}(\sigmabold)$, respectively, and $\overline{O}_{ij}(\sigmabold)$ is the complement of
$O_{ij}(\sigmabold)$ with respect to $\TboldNrij$.

Equation~(\ref{phi_oij_final}) has been expressed in a way that makes it straightforward to explain and
convert into an algorithm for calculation, as I attempt to explain now.
The expression can be described in the following terms: i) separate the hyperedges from $\TboldNr$
into two groups, one that can influence $ij$ over all possible configurations, namely $\TboldNrij$, 
and another that cannot ($\TbarboldNrij$), ii) identify out of $\TboldNrij$ the hyperedges of $\sigmabold$ 
visiting $i$ and $j$, $V_{ij}(\sigmabold)$, iii) only when  
$|V_{ij}(\sigmabold)|\equiv |O_{ij}(\sigmabold)|=o_{ij}$, $\sigmabold$ contributes to $\phi_{ij}(o_{ij},\pbold)$, 
and iv) since there are numerous ways to choose $O_{ij}(\sigmabold)$ from $\TboldNrij$, one requires
the set $\Oboldij$, which contains all those choices, i.e., is the ensemble of allowed configurations.
Consequently, the last line of Eq.~(\ref{phi_oij_final}) can be read as the sum of probabilities over all
possible configurations $\Oboldij$ of hyperedge sets $O_{ij}(\sigmabold)$, where each hyperedge belongs 
to $\TboldNrij$.
Note that there are $|\TboldNrij|={N-2\choose r-2}$ hyperedges to choose from and 
each $O_{ij}(\sigmabold)$ (configuration $\sigmabold$) picks $o_{ij}$ of them, and therefore 
$|\Oboldij|={{N-2\choose r-2}\choose o_{ij}}$.
It is worth mentioning that $\sigmabold$ is used in $O_{ij}(\sigmabold)$ to emphasize its origin
as a particular hypergraph configuration, but that it becomes redundant when the meaning
of $\Oboldij$ is fixed as the collection of configurations (the ensemble) contributing to 
$\phi_{ij}(o_{ij},\pbold)$; at this point, each $O_{ij}$ specifies a unique configuration and no further
reference to $\sigmabold$ is necessary. 

In fact, dropping $\sigmabold$ from $O_{ij}$ offers a combinatorial picture for the last
line of Eq.~(\ref{phi_oij_final}) and other distributions in this section. 
Since each $\sigmabold$ 
is a set of hyperedges connecting non-repeated nodes in cliques of rank $r$, one can think
of each hyperedge as an $r$-tuple of non-repeated indices taken from $\{1,\dots,N\}$, and a
configuration $\sigmabold$ as a collection of non-repeated $r$-tuples. Therefore, $\TboldNr$ is the collection
of all possible $r$-tuples, $\TboldNrij$ the subset of $\TboldNr$ containing all $r$-tuples
that have indices $i$ and $j$ simultaneously, each $O_{ij}$ a sample without replacement of
$o_{ij}$ $r$-tuples taken from $\TboldNrij$, and $\Oboldij$ the collection of all possible
samplings. This way to think about Eq.~(\ref{phi_oij_final}) transfers the emphasis from a graph theoretic problem
to a purely combinatorial one. The cardinalities of all the sets calculated before follow naturally.

The average of $o_{ij}$ can be determined from Eq.~(\ref{phi_oij_final}), through 
$\langle o_{ij}(\pbold)\rangle=\sum_{o_{ij}=0}^{N-2\choose r-2} o_{ij}\phi(o_{ij},\pbold)$, or by calculating 
$\sum_{\sigmaconf}o_{ij}(\sigmabold)P(\sigmabold)$. The result is
\begin{equation}
\langle o_{ij}(\pbold)\rangle=\sum_{(i_1,\dots,i_r)\in\TboldNrij}p_{i_1,\dots,i_r}
\label{avg_oij_final}
\end{equation}
which fits intuition, stating that the expectation of the number of hyperedges visiting the pair $ij$, is
the sum of expectations of each hyperedge that can visit $ij$ to be present over the ensemble.

\begin{figure}
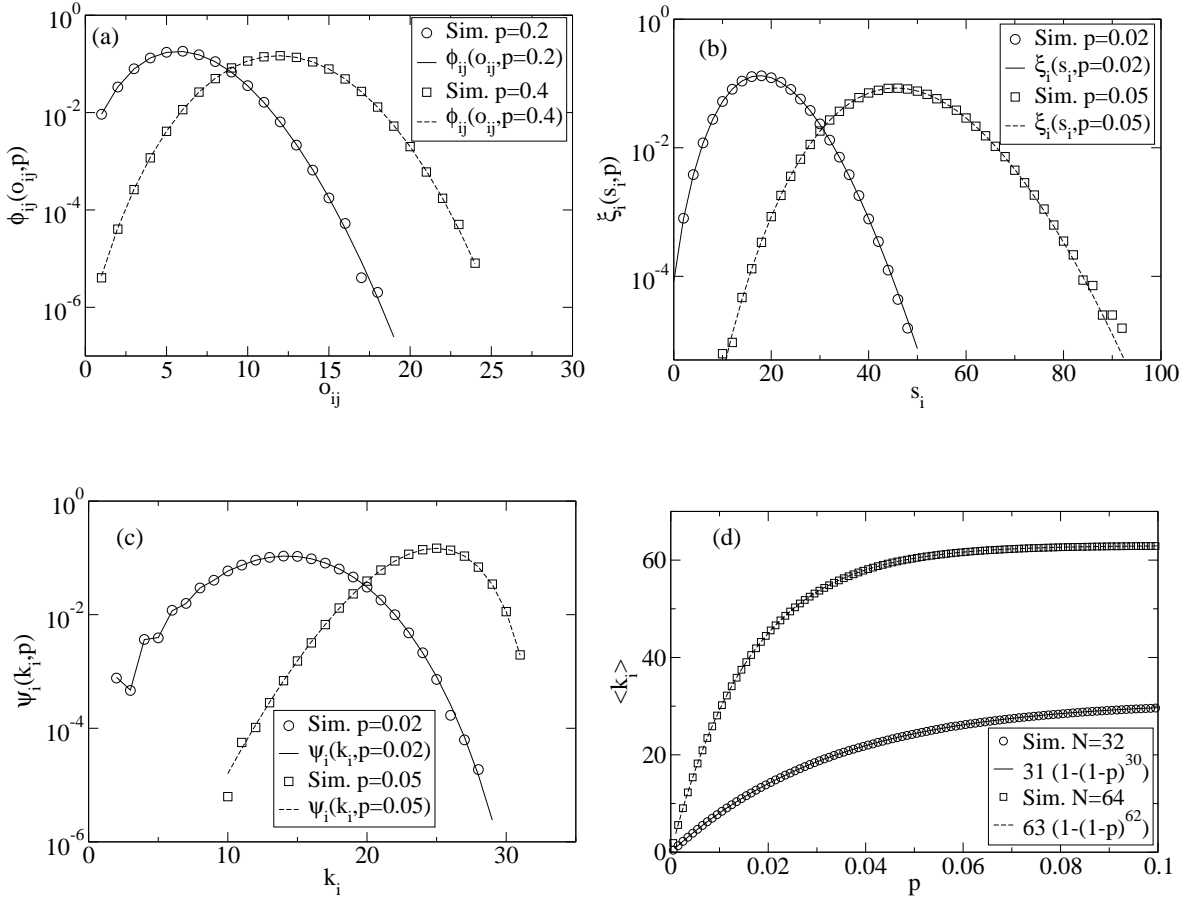

\epsfig{file=fig2a-v2.eps,scale=0.3}
\epsfig{file=fig2b-v2.eps,scale=0.3}\\
\vspace{1cm}
\epsfig{file=fig2c-v2.eps,scale=0.3}\hspace{0.3cm}
\epsfig{file=fig2d-v2.eps,scale=0.3}
\caption{Comparison between theoretical distributions (lines) and simulations (symbols) for
distributions of homogeneous projected networks for $N=32$ and $r=3$: 
(a) $\phi_{ij}(o_{ij},p)$ from Eq.~(\ref{phi_oij_homo_final}) for $N=32$ and corresponding simulations 
($\bigcirc$ for $p=0.2$ and $\square$ for $p=0.4$);
(b) $\xi_i(s_i,p)$ from Eq.~(\ref{xi_si_homo_final}) for $N=32$ and corresponding simulation ($\bigcirc$ for $p=0.02$
and $\square$ for $p=0.05$);
(c) $\psi_i(k_i,p)$ from Eq.~(\ref{psi_ki_homo_final}) for $N=32$ and corresponding simulations ($\bigcirc$ for 
$p=0.02$ and $\square$ for $p=0.05$). 
(d) Average degree $\langle k_i\rangle$ as a function of $p$ in homogeneous networks from 
Eq.~(\ref{avg_ki_homo_final}) and from simulations ($\bigcirc$ for $N=32$ and $\square$ for $N=64$).}
\label{distributions}
\end{figure}

In the homogeneous case, where $p_{i_1,\dots,i_r}=p$ for all $(i_1,\dots,i_r)$, $\phi_{ij}(o_{ij},\pbold)
\to\phi_{ij}(o_{ij},p)$ becomes
\begin{equation}
\phi_{ij}(o_{ij},p)={{N-2\choose r-2}\choose o_{ij}}p^{o_{ij}}(1-p)^{{N-2\choose r-2}-o_{ij}},
\label{phi_oij_homo_final}
\end{equation}
a binomial distribution with $\langle o_{ij}\rangle={N-2\choose r-2}p$ (Fig.~\ref{distributions}(a)). 
Both this average, and more generally Eq.~(\ref{avg_oij_final}), have an interesting 
interpretation explained below regarding signatures of multiway
interactions in observational studies. Another noteworthy fact exhibited by
Eq.~(\ref{phi_oij_homo_final}), even in this very simple case of homogeneous $p$, 
is that $o_{ij}$ does not have a fixed value but instead follows a probability distribution
consequence of the projection process.

Some general features of $\mathcal{P}$ can now be described if one conditions it to be
a monotonic smooth projection, satisfying the inverse function theorem. This condition offers a way to
formally write the distribution of $w_{ij}$ from the distribution of $o_{ij}$ because there is 
a one-to-one relation between the two quantities. 
Furthermore, one can make use of the monotonicity in both the discrete and continuous variable cases.
For the discrete case, defining the distribution $\mu_{ij}(w_{ij},\pbold)$ of weights,
it is simple to see that $\mu_{ij}(w_{ij},\pbold)=\phi_{ij}(\mathcal{P}(o_{ij}),\pbold)$
and the set of possible $w_{ij}$ is obtained by applying Eq.~(\ref{wij-P}) to the domain of $o_{ij}$.
In the continuous case, introducing the densities 
$\mu_{ij}(w_{ij},\pbold)\to\tilde{\mu}_{ij}(w_{ij},\pbold)dw_{ij}$
and $\phi_{ij}(o_{ij},\pbold)\to\tilde{\phi}_{ij}(o_{ij},\pbold)do_{ij}$,
the change of variables theorem for probability distributions implies 
\begin{equation}
\tilde{\mu}_{ij}(w_{ij},\pbold)=\frac{\tilde{\phi}_{ij}(\mathcal{P}^{-1}(w_{ij}),\pbold)}
{\mathcal{P}'(\mathcal{P}^{-1}(w_{ij}))},
\end{equation}
where $\mathcal{P}'$ is the derivative of $\mathcal{P}$. The additive projection $\mathcal{P}_a$
satisfies monotonicity in a trivial way because it is just the identity function. However,
a large class of functions also satisfy monotonicity, including all power law and logarithmic
growth functions.
The nominal projection, on the other hand, does not satisfy the condition because any value of
$o_{ij}\geq 1$ leads to the same weight $w_{ij}=1$, and thus the inverse of $\mathcal{P}_n$ is
not uniquely defined. Regarding the influence of $\tilde{\phi}_{ij}$, if this distribution is sufficiently narrow 
in comparison to the shape of $\mathcal{P}$, 
asymptotic estimates of $\tilde{\mu}_{ij}(w_{ij},p)$ and
it properties (e.g., moments) can be straightforwardly obtained.

As I now describe, $\langle o_{ij}(\pbold)\rangle$ carries a signature of multiway interactions
($r>2$) that can emerge in a large class of weighted projected networks;
if such signature is observed in independently collected weighted network data for which
there is no direct observation of multiway interactions, it would be reasonable
to suspect the presence of these interactions hidden underneath the weighted network.
To explain this, consider Eq.~(\ref{avg_oij_final}) for the ensemble of hypergraphs 
(although the following results are also valid for a single typical large enough hypergraph) 
and label $\pbold_{ij}$ the set of probabilities
$\{p_{i_1,\dots,i_r}|(i_1,\dots,i_r)\in\TboldNrij\}$ of the hyperedges that visit $ij$,
with $\overline{p}_{ij}$ the average and $\Delta p_{ij}$ the variance over the $\pbold_{ij}$ elements. 
For simplicity, let us assume that $N$ increases to $N+\delta N$, and each new node $N+1,N+2,\dots,N+\delta N$ 
connects to $ij$ via new hyperedges. Thus, e.g. for nodes $i,j$, and $N+1$ there are ${N+\delta N-3\choose r-3}$ 
new hyperedges which can potentially occur in a hypergraph, and similarly for all other new nodes. 
Let us also assume that the total set of hyperedges 
$\pbold_{ij}+\boldsymbol{\delta}\pbold_{ij}=\{p_{i_1,\dots,i_r}|(i_1,\dots,i_r)\in\boldsymbol{T}_{ij}(N+\delta N,r)\}$ 
after the addition of $\delta N$ has the same average as $\pbold_{ij}$, i.e., 
$\overline{p_{ij}+\delta p_{ij}}=\overline{p}_{ij}$ and variance $\Delta(p_{ij}+\delta p_{ij})=\Delta p_{ij}$; 
this is satisfied if the $p_{i_1,\dots,i_r}$ in our model are all drawn from the same distribution. Both
$\overline{p}_{ij}$ and $\Delta p_{ij}$ are finite because $0\leq p_{i_1,\dots,i_r}\leq 1$.
Under these conditions, the central limit theorem~\cite{Feller} 
applies to the elements of $\pbold_{ij}+\boldsymbol{\delta}\pbold_{ij}$ and $\pbold_{ij}$.
Hence, $\langle o_{ij}(\pbold+\boldsymbol{\delta}\pbold)\rangle\to {N+\delta N-2\choose r-2}\overline{p}_{ij}$ 
asymptotically in $N$. In fact, if every (large enough) subset of $\pbold_{ij}$ is such that $\overline{p}_{ij}$
and $\Delta p_{ij}$ are constant, $\langle o_{ij}(\pbold)\rangle\to {N-2\choose r-2}\overline{p}_{ij}$.
The set of conditions stipulated above are naturally attained in the heterogeneous hypergraph model, 
where the elements of $\pbold_{ij}$ can be looked at as independent random variables.
The homogeneous case, with all $p_{i_1,\dots,i_r}=p$, satisfies this average exactly even away from the large $N$ limit
as seen from the average of Eq.~(\ref{phi_oij_homo_final}).

The relevance of this result lies in the fact that there is a qualitative change in the behavior
of $\langle o_{ij}(\pbold)\rangle$ for $r=2$ and $r>2$. In the former
$\langle o_{ij}(\pbold)\rangle\to\overline{p}_{ij}$ which does not scale with $N$, but in the later it scales as
$\overline{p}_{ij} N^{r-2}/(r-2)!$~\cite{fn-other-quantities}; i.e., $r>2$ makes $\langle o_{ij}(\pbold)\rangle$
monotonically increasing with $N$. 
To appreciate the importance of this, consider a weighted network 
where one observes that link weights $\langle w_{ij}\rangle$ and size $N$ evolve together with positive correlation
(say, both $N$ and the set of $\langle w_{ij}\rangle$ grow together). 
Two possible qualitative pictures come to mind to explain this behavior: i) the underlying
system is in fact composed of a number of multiway interactions that superimpose
to produce a weighted network, roughly as described in this article, or ii) the system is
pairwise, but the addition of nodes not only introduces
edges between each new node and an old existing node, but also the existing nodes tend to strengthen the
connections between each other by some additional mechanism.
The later model, although possible, is much less natural, and represents
a less economical explanation of the correlation between $N$ and link weights. 
Therefore, being conservative in our interpretation, if $N$ is positively correlated with 
some or all of the $\langle w_{ij}\rangle$, this should be taken as a first indication
that multiway interactions can be present.
Note that the absence of correlations between $N$ and $\langle w_{ij}\rangle$ does
not exclude the presence of multiway interactions, as specific cases may have additional
effects not covered here that may obscure the signatures, e.g., local
multiway relations, interactions between hyperedges, etc.

Certainly, one could argue that for there
to be a weighted network with the behavior described here, it is also necessary to impose
conditions on the type of projection that applies. This is indeed true, but the conditions
necessary to obtain correlation between $\langle w_{ij}\rangle$ and $N$ are relatively modest and well
justified in numerous circumstances. For instance, here I focus on a monotonically
increasing $\mathcal{P}$, which proves sufficient. Furthermore, it is even acceptable to have
a $\overline{p}_{ij}$ that decreases with $N$, just as long as the decrease is slower than
$N^{-(r-2)}$. The level of detail known about $\mathcal{P}$ and $\pbold_{ij}$ goes
hand in hand with the detail that can be learned about the multiway interactions.
For instance, if $\overline{p}_{ij}=p$ (homogeneous) and $\mathcal{P}=\mathcal{P}_a$,
then $\ln\langle w_{ij}\rangle$ vs. $\ln N$ would yield the value of $r$. If, on the other hand, 
all one knows is that $\mathcal{P}$ is monotonically increasing, the $\ln\langle w_{ij}\rangle$ vs. $\ln N$
would offer an estimate of the combined effects of $\mathcal{P}$ and $\pbold_{ij}$.
Overall, the present discussion suggest that {\it in the case of evolving networks
with correlations between $N$ and the set of $\langle w_{ij}\rangle$, it is reasonable to suspect
multiway interactions active in the background, and further exploration for evidence of such interactions
is well justified.} These results are general, and do not
need to be specialized into a particular example. However, if one is interested in
using the results of this article as a method to attempt to determine quantitative details of the multiway
interactions that may be present, additional work is needed to extend the results to more
detailed and perhaps slightly different situations.

The two projections $\mathcal{P}_a$ and $\mathcal{P}_n$ can now be explained further. 
For $\mathcal{P}_a$, the properties of $w_{ij}$ are those of $o_{ij}$, and thus already calculated. 
The other property to describe is the so-called strength $s_i$ of node $i$, 
equal to $\sum_j o_{ij}$.
It is intuitively helpful to calculate the distribution
of strengths $\xi_i(s_i,\pbold)$ by making use of the relation between $s_i$ and
$\ell_i$, the number of hyperedges visiting $i$. These two quantities relate via $s_i=(r-1)\ell_i$, 
and one can determine the distribution $\zeta_i(\ell_i,\pbold)$ of $\ell_i$ and from it
compute $\xi_i(s_i,\pbold)$.
Note that while $s_i$ is a property of the graph, $\ell_i$ is a property of the hypergraph. 
Once again, the independence of the components of $\pbold$ simplifies the sum over configurations $\sigmaconf$
(notation in Table~\ref{table-zeta_elli}).
The hyperedges that could affect $\ell_i$ belong to $\TboldNri$, the collection of all
hyperedges visiting $i$ in $\TboldNr$, 
and $L_i(\sigmabold)$ is the set of hyperedges from $\TboldNri$ in configuration
$\sigmabold$ (when $|L_i(\sigmabold)|=\ell_i$ one writes it as $\lambda_i(\sigmabold)$). 
From the definition $\zeta_i(\ell_i,\pbold)=\sum_{\sigmaconf}
\delta(\ell_i,\sum_{(i_1,\dots,i_r)\in L_i(\sigmabold)}\sigma_{i_1,\dots,i_r})P(\sigmabold,\pbold)$, 
one can quickly conclude that
\begin{equation}
\zeta_i(\ell_i,\pbold)=\sum_{\lambda_i(\sigmabold)\in\boldsymbol{\Lambda}_i}
\prod_{(i_1,\dots,i_r)\in\lambda_i(\sigmabold)}p_{i_1,\dots,i_r}
\prod_{(i_1,\dots,i_r)\in\overline{\lambda}_i(\sigmabold)}(1-p_{i_1,\dots,i_r}),
\label{zeta_elli_final}
\end{equation}
where $\boldsymbol{\Lambda}_i$ is the ensemble of configurations $\lambda_i(\sigmabold)$,
and $\lambda_i(\sigmabold)\bigcup\overline{\lambda}_i(\sigmabold)=\TboldNri$. 
Then 
\begin{equation}
\xi_i(s_i,\pbold)=\zeta_i(s_i/(r-1),\pbold)/(r-1),
\label{xi_si_final}
\end{equation}
where $s_i$ takes values $0,r-1, 2(r-1), \dots,(r-1){N-1\choose r-1}$.
Once again, an equivalence between hyperedge sets and combinatorics can be drawn: $\TboldNri$ is the union of
all $r$-tuples drawn from $\{1,\dots,N\}$ with one element always $i$, and thus there are
$|\TboldNri|={N-1\choose r-1}$ $r$-tuples in total. 
Each $\lambda_i$ is a distinct choice of $\ell_i$ of these $r$-tuples; 
clearly $|\boldsymbol{\Lambda}_i|={{N-1\choose r-1}\choose\ell_i}$. The sum 
$\sum_{\lambda_i\in\boldsymbol{\Lambda}_i}$
is a sum over all choices of $\ell_i$ $r$-tuples from $\TboldNri$.
The averages of these quantities are given by
\begin{equation}
\langle\ell_i(\pbold)\rangle=\sum_{(i_1,\dots,i_r)\in\TboldNri}p_{i_1,\dots,i_r}
\label{avg_elli_final}
\end{equation}
and 
\begin{equation}
\langle s_i(\pbold)\rangle=(r-1)\langle\ell_i(\pbold)\rangle
=(r-1)\sum_{(i_1,\dots,i_r)\in\TboldNri}p_{i_1,\dots,i_r}.
\label{avg_si_final}
\end{equation}
For the homogeneous case, 
$\zeta_i(\ell_i,p)={{N-1\choose r-1}\choose\ell_i}p^{\ell_i}(1-p)^{{N-1\choose r-1}-\ell_i}$,
with average $\langle\ell_i\rangle={N-1\choose r-1}p$. Therefore,
\begin{equation}
\xi_i(s_i,p)={{N-1\choose r-1}\choose s_i/(r-1)}p^{s_i/(r-1)}(1-p)^{{N-1\choose r-1}-s_i/(r-1)},
\label{xi_si_homo_final}
\end{equation}
and $\langle s_i\rangle=(r-1){N-1\choose r-1}p$ (see Fig.~\ref{distributions}(b)).

The nominal interaction $\mathcal{P}_n$ needs a different treatment. Note that under
this projection, $w_{ij}$ can be either 0 or 1. To determine the probability for $w_{ij}$, 
$\pi_{ij}(w_{ij},\pbold)$, one merely needs to determine the probabilities that $o_{ij}$ is either
0 or $\geq 1$, that is $\pi_{ij}(w_{ij},\pbold)=\phi_{ij}(o_{ij}=0,\pbold)$ or
$\pi_{ij}(w_{ij}=1,\pbold)=1-\phi_{ij}(o_{ij}=0,\pbold)$. Therefore,
\begin{equation}
\pi_{ij}(w_{ij},\pbold)
=\left\{
\begin{array}{ll}
1-\prod_{(i_1,\dots,i_r)\in\TboldNrij}(1-p_{i_1,\dots,i_r});&\qquad w_{ij}=1\\
\prod_{(i_1,\dots,i_r)\in\TboldNrij}(1-p_{i_1,\dots,i_r});&\qquad w_{ij}=0.
\end{array}
\right.
\label{pi_wij_final}
\end{equation}
In the homogeneous case,
\begin{equation}
\pi_{ij}(w_{ij},p)
=\left\{
\begin{array}{ll}
1-(1-p)^{N-2\choose r-2};&\qquad w_{ij}=1\\
(1-p)^{N-2\choose r-2};&\qquad w_{ij}=0.
\end{array}
\label{pi_wij_homo_final}
\right.
\end{equation}
Both Eqs.~(\ref{pi_wij_final}) and (\ref{pi_wij_homo_final}) are closely related to the average 
number of connections for each node of a projected network, as explained next.

\begin{figure}
\epsfig{file=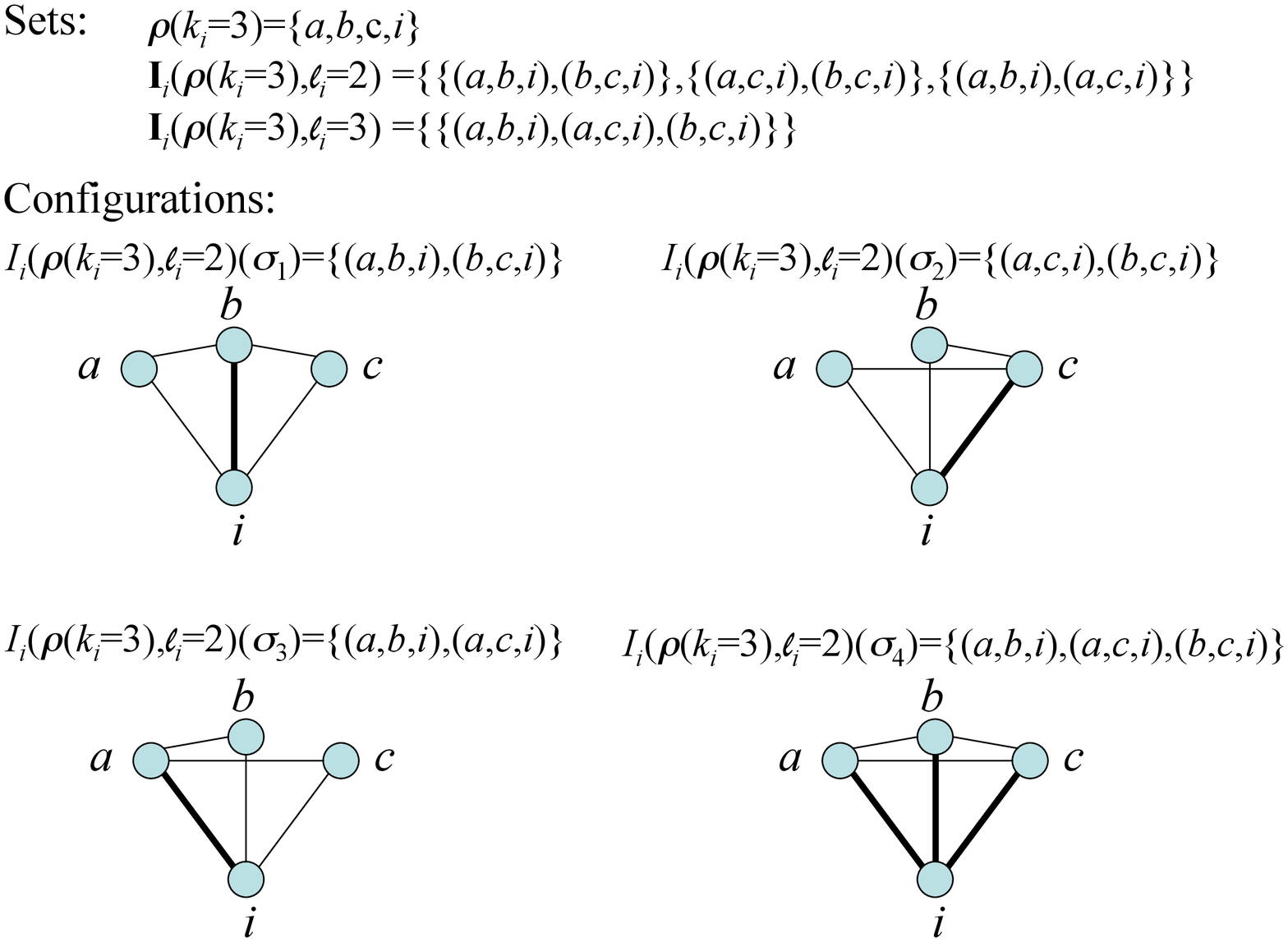,angle=270,scale=0.5}
\caption{Illustration ($r=3$) of the emergence of degree $k_i$ as a consequence of various possible 
hyperedge configurations. The figure also illustrates $Q_{r-1}(k_i,\ell_i)$. 
There are 4 possible ways in which $i$ can be connected to nodes $\{a,b,c\}$, each case corresponding
to one of the configurations shown above ($\sigmabold_1,\sigmabold_2,\sigmabold_3,\sigmabold_4$) 
in the projected network. The sets $\boldsymbol{\rm I}_i(\rho(k_i),\ell_i)$ are defined 
for both $\ell_i=2$ and 3, the only two possible cases. Note also that if one focuses only on the nodes
$\{a,b,c\}$ ignoring $i$, all configurations can be mapped to the construction of all possible cliques
of size 2 of these nodes, generating $Q_{r-1=2}(k_i=3,\ell_i=2)=3$ and $Q_{r-1=2}(k_i=3,\ell_i=3)=1$. 
The fact that all configurations are globally 
connected is an accident due to the small value of $k_i=3$, but in general, nodes simply 
need to belong to $\ell_i$ cliques of size $r-1$. Finally, note the thickness of links, representative
of $o_{ij}$.}
\label{ki-illustration}
\end{figure}

For any weighted projected network generated from a $\mathcal{P}$ satisfying $o_{ij}>0\Rightarrow w_{ij}>0$ and
$o_{ij}=0\Rightarrow w_{ij}=0$, such as $\mathcal{P}_a$ and $\mathcal{P}_n$, the number of 
connections $k_i$ visiting node $i$ are characterized by $\psi_i(k_i,\pbold)$, the distribution of $k_i$
(an expanded and pedagogical exposition of the calculation and its consequences can be found in~\cite{Lopez-pk}). 
The degree can be either 0 or take any value from $r-1\leq k_i\leq N-1$.
To determine $\psi_i(k_i,\pbold)$ (notation in Table~\ref{table-psi_ki}), 
one can proceed in a similar way as before: in configuration
$\sigmabold$, the set of hyperedges visiting $i$ and producing degree $k_i$ is $K_i(\sigmabold)$.
This means that hyperedges in $K_i(\sigmabold)$ visit exactly $k_i$ nodes and node $i$.
It is interesting to note that another configuration $\sigmabold'$, associated with $K_i(\sigmabold')$, 
with a different set and/or number of hyperedges can lead to the same $k_i$, because these hyperedges 
still visit the same number of nodes $k_i$ (see Fig.~\ref{ki-illustration} for an illustration). 
With this definition, one can write
\begin{equation}
\psi_i(k_i,\pbold)=\sum_{K_i(\sigmabold)\in\boldsymbol{\rm K}_i}
\prod_{(i_1,\dots,i_r)\in K_i(\sigmabold)}p_{i_1,\dots,i_r}
\prod_{(i_1,\dots,i_r)\in\overline{K}_i(\sigmabold)}(1-p_{i_1,\dots,i_r}),
\end{equation}
where $\boldsymbol{\rm K}_i$ is the union of all possible sets $K_i(\sigmabold)$, and
the complement set $\overline{K}_i(\sigmabold)$ satisfies
$K_i(\sigmabold)\bigcup \overline{K}_i(\sigmabold)=\TboldNri$. 
Since the number of hyperedges is not fixed across members of $\boldsymbol{\rm K}_i$, one can
further organize the $K_i(\sigmabold)$ by their numbers of hyperedges $\ell_i(\sigmabold)$.
The bounds of $\ell_i$ are dictated by the following: for degree $k_i$, a minimum of 
$\lceil k_i/(r-1)\rceil$ hyperedges is required ($\lceil .\rceil$ represents the ceiling function), 
and there can be no more than ${k_i \choose r-1}$ hyperedges.
Using this organization, and introducing the notation $K^{(\ell_i)}_i(\sigmabold)$ and 
$\boldsymbol{\rm K}_i(\ell_i)$ to represent, respectively, the sets $K_i(\sigmabold)$ involving 
exactly $\ell_i$ hyperedges and their unions, one can write
\begin{equation}
\psi_i(k_i,\pbold)=\sum_{\ell_i=\left\lceil \frac{k_i}{r-1}\right\rceil}^{k_i\choose r-1}
\sum_{K^{(\ell_i)}_i(\sigmabold)\in\boldsymbol{\rm K}_i(\ell_i)}
\prod_{(i_1,\dots,i_r)\in K^{(\ell_i)}_i(\sigmabold)}p_{i_1,\dots,i_r}
\prod_{(i_1,\dots,i_r)\in\overline{K}^{(\ell_i)}_i(\sigmabold)}(1-p_{i_1,\dots,i_r}).
\end{equation}
The sets $\boldsymbol{\rm K}_i(\ell_i)$ are only subsets of $\boldsymbol{\Lambda}_i$ in which the
$\ell_i$ hyperedges involve exactly $i$ and $k_i$ other nodes. Finally, it is possible
to exploit one more symmetry that facilitates an algorithmic understanding of $\psi_i(k_i,\pbold)$: 
the sets that make up $\boldsymbol{\rm K}_i(\ell_i)$ involve
several possible distinct node sets. However, one can further segregate these sets by
the specific nodes in them. Hence, if one takes a set, $\rho(k_i)$, of $k_i$ specific nodes and $i$, 
there are several configurations in which their associated $K^{(\ell_i)}_i(\sigmabold)$ contain 
$\ell_i$ hyperedges visiting only those nodes. Thus, a configuration with specific $\rho(k_i)$ nodes 
connected to $i$, using $\ell_i$ hyperedges is labelled $I^{(\rho(k_i),\ell_i)}_i(\sigmabold)$, and the
union of configurations is labelled $\boldsymbol{\rm I}_i(\rho(k_i),\ell_i)$.
The union of all sets $\boldsymbol{\rm I}_i(\rho(k_i),\ell_i)$ (which are non-intersecting)
is equal to $\boldsymbol{\rm K}_i(\ell_i)$. This leads to the final expression
\begin{multline}
\psi_i(k_i,\pbold)= \sum_{\rho(k_i)\in \boldsymbol{\rm R}_i(N,k_i)}\\
\sum_{\ell_i=\left\lceil\frac{k_i}{r-1}\right\rceil}^{k_i\choose r-1}
\sum_{I^{(\rho(k_i),\ell_i)}_i(\sigmabold)\in\boldsymbol{\rm I}_i(\rho(k_i),\ell_i)}
\prod_{(i_1,\dots,i_r)\in I^{(\rho(k_i),\ell_i)}_i(\sigmabold)}p_{i_1,\dots,i_r}
\prod_{(i_1,\dots,i_r)\in\overline{I}^{(\rho(k_i),\ell_i)}_i(\sigmabold)}(1-p_{i_1,\dots,i_r}),
\label{psi_ki_final}
\end{multline}
where $\overline{I}^{(\rho(k_i),\ell_i)}_i(\sigmabold)$ is the complement of 
$I^{(\rho(k_i),\ell_i)}_i(\sigmabold)$ with respect to $\TboldNri$, 
and $\boldsymbol{\rm R}_i(N,k_i)$ is the union of all possible $\rho(k_i)$, each one a distinct 
$(k_i+1)$-tuple taken from the set $\{1,\dots,N\}$ with one choice always being $i$. 
The sizes of sets are: $|\boldsymbol{\rm R}_i(N,k_i)|={N-1\choose k_i}$, 
and $|\boldsymbol{\rm I}_i(\rho(k_i),\ell_i)|=Q_{r-1}(k_i,\ell_i)$; the later is the result of 
a combinatorial
problem that can be defined in terms of general graph theory. Specifically, $Q_{r-1}(k_i,\ell_i)$ 
corresponds to the number of distinct hypergraphs that can be constructed with $k_i$ nodes and
$\ell_i$ hyperedges of rank $r-1$, and each node belongs to at least one of the hyperedges~\cite{Lopez-pk}. 
In fact, each $I^{(\rho(k_i),\ell_i)}_i(\sigmabold)$
can be mapped to each one of these hypergraphs. To determine $\langle k_i(\pbold)\rangle$, it is convenient to
use the relation 
\begin{equation}
\langle k_i(\pbold)\rangle=\sum_{\sigmaconf}k_i(\sigmabold)P(\sigmabold)
=\sum_{\sigmaconf}\sum_{j=1;j\neq i}^N\theta(o_{ij}(\sigmabold))P(\sigmabold)
=\sum_{j=1;j\neq i}^N\sum_{\sigmaconf}\theta(o_{ij}(\sigmabold))P(\sigmabold).
\end{equation}
By first summing over a single $j$, one notices that only hyperedges in $\TboldNrij$ must be considered
in detail. Given that
\begin{equation}
\theta(o_{ij}(\sigmabold))=\theta\left(\sum_{(i_1,\dots,i_r)\in O_{ij}(\sigmabold)}\sigma_{i_1,\dots,i_r}\right)
=1-\delta\left(\sum_{(i_1,\dots,i_r)\in O_{ij}(\sigmabold)}\sigma_{i_1,\dots,i_r},0\right),
\end{equation}
one arrives at
\begin{equation}
\langle k_i(\pbold)\rangle=\sum_{j=1;j\neq i}^N
\left[1-\prod_{(i_1,\dots,i_r)\in\TboldNrij}(1-p_{i_1,\dots,i_r})\right].
\label{avg_ki_final}
\end{equation}
When compared with Eq.~(\ref{pi_wij_final}), it becomes evident that each link $ij$ contributes to 
$k_i$ independently.

In the homogeneous case, making use of the combinatorial results presented, one obtains 
(Fig.~\ref{distributions}(c))
\begin{equation}
\psi_i(k_i,p)={N-1\choose k_i}\sum_{\ell_i=\left\lceil\frac{k_i}{r-1}\right\rceil}^{k_i\choose r-1}
Q_{r-1}(k_i,\ell_i)p^{\ell_i}(1-p)^{{N-1\choose r-1}-\ell_i}.
\label{psi_ki_homo_final}
\end{equation}
Without diving into too much detail, $Q_{r-1}(k_i,\ell_i)$ can be calculated via the inclusion-exclusion
principle of combinatorics~\cite{Riordan,Lopez-pk}, which produces
\begin{equation}
Q_{r-1}(k_i,\ell_i)=\sum_{m=0}^{k_i}(-1)^{k_i-m}{k_i\choose m}{{m\choose r-1}\choose\ell_i}.
\end{equation}
Among the identities satisfied by $Q_{r-1}(k,\ell)$~\cite{Lopez-pk}, one finds that 
${{N-1\choose r-1}\choose \ell}=\sum_{k} {N-1\choose k}Q_{r-1}(k,\ell)$, which is used to show
normalization of $\psi_i(k_i,p)$. Another identity, 
$\sum_k k{N-1\choose k}Q_{r-1}(k,\ell)=
(N-1)\left[{{N-1\choose r-1}\choose\ell}-{{N-2\choose r-1}\choose\ell}\right]$, leads to
the average of $\psi_i(k_i,p)$, 
\begin{equation}
\langle k_i\rangle=(N-1)\left[1-(1-p)^{N-2\choose r-2}\right],
\label{avg_ki_homo_final}
\end{equation}
where the brackets are equal to $\pi_{ij}(w_{ij}=1,p)$ from Eq.~(\ref{pi_wij_homo_final}) 
(see Fig.~\ref{distributions}(d)). This average can also be calculated directly from 
Eq.~(\ref{avg_ki_final})~\cite{Lopez-pk}.

To conclude this section, it is useful to point out how the previous results can be connected with
concrete problems. The logic is similar to that found in \cite{Park,Bradde}, in which
the ensemble is chosen to fit observations. In the framework presented here, it is possible to choose 
the hypergraph ensemble to fit hypergraph properties (such as Eqs.~(\ref{avg_oij_final}) or (\ref{avg_elli_final})),
projected network properties (Eqs.~(\ref{avg_si_final}) or (\ref{avg_ki_final})), or a combination of
both (as long as it is well defined); the choice comes down to practical considerations such as the
available data one intends to fit, or the belief that certain mechanisms may be at play and 
therefore must be part of the model.
Once an ensemble is defined (satisfying the assumptions of hyperedges which are non-interacting, 
undirected, and with uniform rank), the expressions derived above for the heterogeneous ensemble apply, 
but an additional set of constraints emerges for the $p_{i_1,\dots,i_r}$ guaranteeing that the entropy 
is maximized, distinguishing the situation from that of the fully heterogeneous ensemble, where 
each $p_{i_1,\dots,i_r}$ is free to have any value between 0 and 1.

As an example, consider the ensemble that specifies strengths $\langle s_i\rangle$ on the projected 
networks with projection $\mathcal{P}_a$. This can be constructed from the Hamiltonian
\begin{equation}
H(\sigmabold)=\sum_{i=1}^N\beta_i s_i(\sigmabold)
=(r-1)\sum_{(i_1,\dots,i_r)\in\sigmabold}(\beta_{i_1}+\dots+\beta_{i_r})\sigma_{i_1,\dots,i_r}.
\end{equation}
This ensemble is 
completely specified by calculating the relation between $\langle\sigma_{i_1,\dots,i_r}\rangle$,
by definition equal to $p_{i_1,\dots,i_r}$, and the set of parameters $\{\beta_1,\dots,\beta_N\}$.
After determining $P(\sigmabold)$, one can 
compute $\langle\sigma_{i_1,\dots,i_r}\rangle=\sum_{\sigmaconf}\sigma_{i_1,\dots,i_r}P(\sigmabold)$
to find
\begin{equation}
\langle\sigma_{i_1,\dots,i_r}\rangle=p_{i_1,\dots,i_r}
=\frac{e^{-(r-1)(\beta_{i_1}+\dots+\beta_{i_r})}}{1+ e^{-(r-1)(\beta_{i_1}+\dots+\beta_{i_r})}}
\end{equation}
where the parameters satisfy Eq.~(\ref{avg_si_final}), and therefore
\begin{equation}
\langle s_i\rangle=(r-1)\sum_{(i_1,\dots,i_r)\in\TboldNri}p_{i_1,\dots,i_r}
=(r-1)\sum_{(i_1,\dots,i_r)\in\TboldNri}\frac{e^{-(r-1)(\beta_{i_1}+\dots+\beta_{i_r})}}
{1+ e^{-(r-1)(\beta_{i_1}+\dots+\beta_{i_r})}}.
\end{equation}
One way to understand this result is from the relation
\begin{equation}
dp_{i_1,\dots,i_r}=-(r-1)\sum_{g=1}^r\frac{p_{i_1,\dots,i_r}}{1-p_{i_1,\dots,i_r}}d\beta_{i_g}.
\end{equation}
If only $\beta_{i_g}$ changes (by, say, $d\beta_{i_g}$), hyperedges without node $i_g$ are unaffected, and 
those with $i_g$ all increase in probability proportionally to $d\beta_{i_g}$. 
As in Ref.~\cite{Park}, the $\beta_i$ can be taken
from a distribution, leading in turn to a distribution of $\langle s_i\rangle$. This can be used
to obtain a desired distribution of $\langle s_i\rangle$ as dictated by the problem.
\section{Percolation properties and sparse cases}
\label{percolation}
Another important aspect of the hypergraph ensemble and its projected networks is their 
percolation properties. To calculate these, one can use the equivalence, first pointed out by
Fortuin and Kasteleyn~\cite{Fortuin}, between percolation and the mean-field $q$-states Potts model 
at $q\to 1$. The solution to the later model consists of determining the state of the nodes,
and whether there is a phase transition.
The solution and its properties can be obtained by studying the model's Helmholtz free energy. A detailed
development of equivalence of the models can be found in Refs.~\cite{Bradde,Engel}; here,
I set up the calculation starting at
the free energy and develop the percolation properties from there. I consider the homogeneous
case only, although it is possible to solve some forms of heterogeneous models. 

Consider the Hamiltonian of the general $q$-state Potts model with $N$ nodes,
$H_q=-\sum_{i_1,\dots,i_r} J_{i_1,\dots,i_r} \delta(u_{i_1},\dots,u_{i_r})$, 
where $u_{i_1},\dots,u_{i_r}$ represent the respective spin states of the nodes $i_1,\dots,i_r$ 
from the possible states $1,\dots,q$, and $J_{i_1,\dots,i_r}$ the strength of the interaction among them.
A hyperedge exists among nodes $i_1,\dots,i_r$ if $u_{i_1}=\dots=u_{i_r}$, i.e., if these nodes are 
in the same spin state.
Let us denote the number of system nodes with spin $u$ as $N_u$, and the density of these 
as $c_u=N_u/N$, which satisfies $\sum_u c_u=1$. In the homogeneous system, 
since $J_{i_1,\dots,i_r}=J$ and given that only
$r$-tuples of equal spins contribute to $H_q$ (i.e., only hyperedges), the energy is equal 
to $H_q=-J\sum_u {N_u\choose r}$, 
the sum of interaction energies among all hyperedges having equal spin. The connection between 
percolation and the Potts model carries with it the relation $J=-\ln(1-p)$, and for small
$p$, this approximates to $J\approx p$.

In order to find the
Helmholtz free energy of the system, one must first determine the partition function $Z_q$. In this
model, it can be written on the basis of all configurations of state values $u_i$, or in terms of
the set of numbers $\{N_u\}_{u=1,\dots,q}$. Using the later set of variables, and taking into account
the multiplicity in the choices for each node state, one arrives at
\begin{equation}
Z_q=\sum_{\sum_{u=1}^q N_u=N}e^{-\left[-J\sum_u {N_u\choose r}\right]}\frac{N!}{\prod_{u=1}^q N_u!}
=\sum_{\sum_{u=1}^q c_u=1}e^{-\left[-J\sum_u {Nc_u\choose r}+N\sum_u c_u\ln c_u\right]},
\end{equation}
where 
the inverse temperature parameter $\beta$ is absorbed into $J$.
In the canonical ensemble, the free energy is given by $F_q=-\ln Z_q$.
When the interaction $J$ is too weak to keep the nodes ordered collectively in groups of common states, 
the solution to the problem is expected to be symmetric, i.e. $c_u=1/q$ (all states are
equally occupied). However, as the interaction strengthens, one would expect that symmetry is
broken and one state (say $u=1$) becomes dominant. By these arguments, $F_q$ can be sought
by introducing the {\it ansatz}
\begin{equation}
c_u=\left\{
\begin{array}{lr}
\frac{1+(q-1)\tilde{f}_q}{q}&\qquad;u=1~\\
\frac{1-\tilde{f}_q}{q}&\quad;u\neq 1,
\end{array}
\right.
\label{c_symmetry_breaking}
\end{equation}
where $\tilde{f}_q$ is the fractional size of the system in state $u=1$, and the condition $\sum_u c_u=1$
is automatically satisfied. This leads to
\begin{equation}
Z_q=\int d\tilde{f}_q e^{\left[J{\frac{N}{q}(1+(q-1)\tilde{f}_q)\choose r}
+J(q-1){\frac{N}{q}(1-\tilde{f}_q)\choose r}
-\frac{N}{q}(1+(q-1)\tilde{f}_q)\ln\frac{1+(q-1)\tilde{f}_q}{q}
-\frac{N}{q}(1-\tilde{f}_q)\ln\frac{1-\tilde{f}_q}{q}\right]}.
\label{Zq_fq}
\end{equation}
In the thermodynamic limit ($N\to\infty$), the Laplace method of integration can be applied to $Z_q$
\cite{Bender}. Once applied, $F_q=-\ln Z_q$ yields to leading order
\begin{multline}
F_q=-J{\frac{N}{q}(1+(q-1)f_q)\choose r}
-J(q-1){\frac{N}{q}(1-f_q)\choose r}\\
+\frac{N}{q}(1+(q-1)f_q)\ln\frac{1+(q-1)f_q}{q}
+\frac{N}{q}(1-f_q)\ln\frac{1-f_q}{q}
\label{Fq_extr}
\end{multline}
where $f_q$ is the value of $\tilde{f}_q$ for which the exponent of the argument of the $Z_q$
integral is maximized.
Explicitly, $f_q$ is obtained by equating to 0 the first derivative with respect to $\tilde{f}_q$ 
of the exponent in Eq.~(\ref{Zq_fq}), and using $c_1(f_q)$ and $c_u(f_q)$ to refer to the 
the fractions $c_u$ from Eq.~(\ref{c_symmetry_breaking}) evaluated at $u=1$ and $u\neq 1$, 
respectively. Thus, $f_q$ must satisfy
\begin{equation}
\ln\left(\frac{1-f_q}{1-(q-1)f_q}\right)
=-J\left[{Nc_1(f_q)\choose r}\sum_{i=1}^{r-1}\frac{1}{Nc_1(f_q)-i}
-{Nc_u(f_q)\choose r}\sum_{i=1}^{r-1}\frac{1}{Nc_u(f_q)-i}\right].
\label{fq_self_consistent}
\end{equation}
This is the self-consistency equation for the fractional size of the component of broken symmetry.
For $q=1$, $f\equiv f_{q=1}$ is the fractional size of the percolating spanning cluster. 
Note that $f_q=0$ is also a solution to Eq.~(\ref{fq_self_consistent}), but its stability
breaks down when the second derivative of the exponent of the integrand of $Z_q$ changes sign. 
The value of $J$ for which the sign change occurs is given by the relation
\begin{equation}
p_c\approx J_c=\left[N\frac{\partial^2}{\partial N^2}{N\choose r}\right]^{-1},
\label{Jc}
\end{equation}
where $q=1$ has already been introduced (otherwise the solution would be the same but with $N/q$
in place of $N$ everywhere).

\begin{figure}
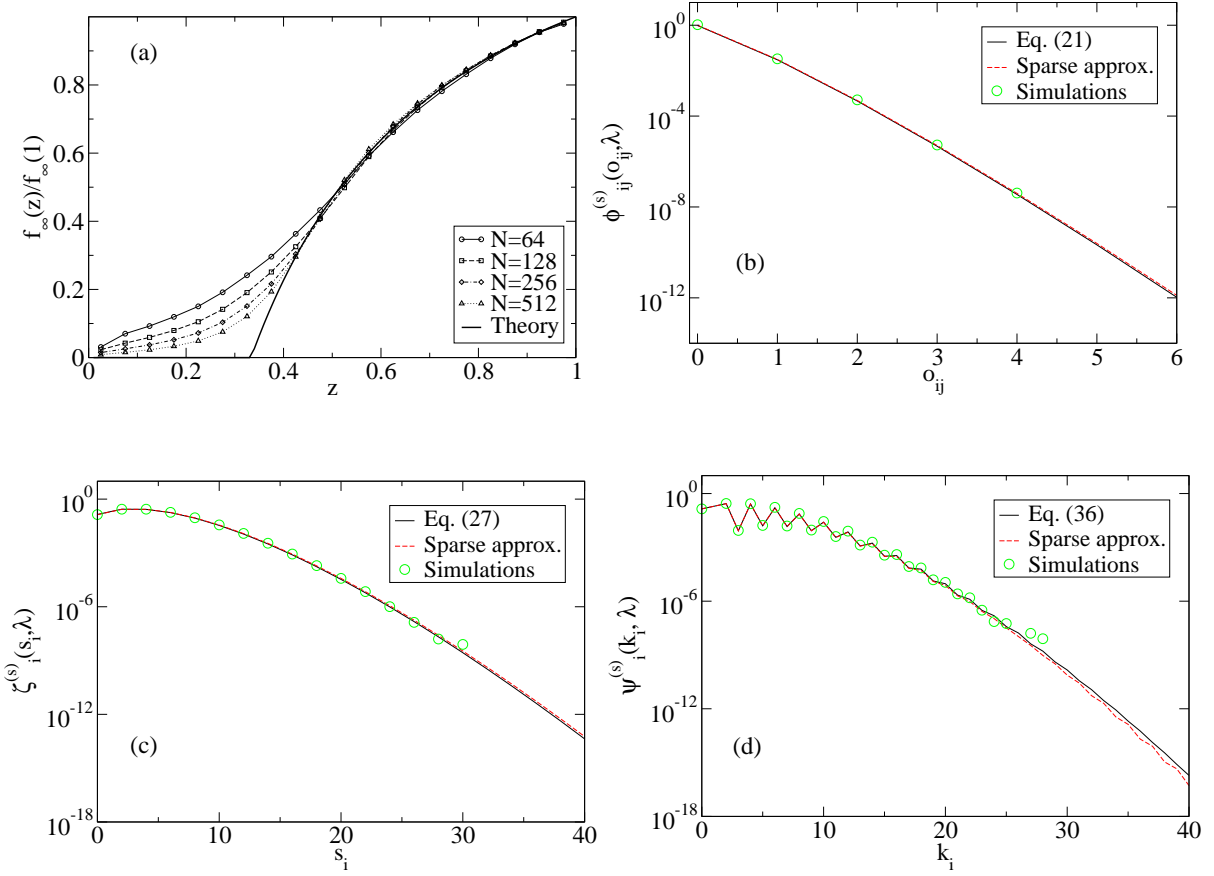

\epsfig{file=fig4a-v2.eps,scale=0.3}
\epsfig{file=fig4b-v2.eps,scale=0.3}\\
\vspace{1cm}
\epsfig{file=fig4c-v2.eps,scale=0.3}
\epsfig{file=fig4d-v2.eps,scale=0.3}
\caption{(color online) Percolation limit for the ensemble of $\langle s\rangle=\lambda_\text{max}$: 
(a) $f(z)/f(z=1)$ vs. $z$ ($\lambda_\text{max}=3.0$) from Eq.~(\ref{f_final}) (line) and simulations of
$N=64\quad(\bigcirc)$, $N=128\quad(\square)$, $N=256\quad(\Diamond)$ and $N=512\quad(\vartriangle)$.
As the system size increases, the theoretical solution is approached. Projected
network properties ($\mathcal{P}_a$ for $s_i$) for $N=128$ in the ensemble of 
$\langle s\rangle=\lambda=4.0$ predicted
by theory (line), their respective sparse approximations (dashed line) and simulations ($\bigcirc$):
(b) $\phi^{(s)}_{ij}(o_{ij},\lambda)$,
(c) $\xi^{(s)}_i(s_{i},\lambda)$, and
(d) $\psi^{(s)}_i(k_i,\lambda)$.}
\label{fig-dilute}
\end{figure}

In the thermodynamic limit, one can derive a compact equation for $f$ and arbitrary $r$.
Both terms in the brackets of Eq.~(\ref{fq_self_consistent}) are 
polynomials emerging from the derivative $\frac{\partial}{\partial x}{x\choose r}$, 
labeled $M(x,r)=\sum_{m=0}^{r-1}a_r(m) x^m$, evaluated at 
$x=Nc_1$ and $Nc_{u}$ (I continue the same shorthand of $c_{u\neq 1}=c_u$).
Subtracting, one obtains 
$M(N_1,r)-M(N_{u\neq 1},r)=\sum_{m=1}^{r-1}a_r(m)(N_1^m-N_u^m)$,
where the coefficient $m=0$ vanishes. It is possible to express the coefficients $a_r(m)$ in terms
of elementary symmetric polynomials~\cite{MacMahon} and binomial coefficients, but the analysis here is restricted to the 
asymptotic limit, and thus only requires the coefficient $a_r(r-1)$, equal to $1/(r-1)!$ as can be 
determined by inspecting $M(x,r)$.
Using the identity $x^m-y^m=(x-y)\sum_{l=0}^{m-1}x^{m-1-l}y^l$, the ansatz~(\ref{c_symmetry_breaking}),
and the self-consistency relation~(\ref{fq_self_consistent}) with $q=1$, one obtains
\begin{equation}
\ln(1-f)=-Jf\sum_{m=1}^{r-1}a_r(m)N^m\sum_{l=0}^{m-1}(1-f)^l
=-J\sum_{m=1}^{r-1}a_r(m)N^m\left(1-(1-f)^m\right).
\end{equation}
Close to percolation, it is justified to write $J=\lambda J_c$, with $\lambda\geq 1$ and $J_c$ from
Eq.~(\ref{Jc}). By L'Hopital's
rule, for the dominant term in $N$, the size of the largest component emerges as
\begin{equation}
\ln(1-f)=-\frac{\lambda}{r-1}(1-(1-f)^{r-1}),
\label{f_final}
\end{equation}
which generalizes expressions for $f$ for $r=2$ and 3 in Refs.~\cite{Newman-clusters,Miller-clusters}
(these authors tackle the percolation question to illustrate ideas different than those explored here).
To test this expression, it is customary to define the percolation problem with respect to a network (or hypergraph)
that is not complete, but instead is already diluted. By defining the rescaling $z=p/p_{\text{max}}$ 
where typically $p_\text{max}\ll 1$, the original undiluted hypergraph is $z=1$, and percolation occurs at
$z_c=p_c/p_\text{max}$, or if using $\lambda_\text{max}$, $z_c=1/\lambda_\text{max}$ (see Fig~\ref{fig-dilute}(a)).

The percolation transition can be shown to be second order by expanding both sides of Eq.~(\ref{f_final}), which
leads to
\begin{equation}
\sum_{g=1}^\infty\frac{f^g}{g}=\frac{\lambda}{r-1}\sum_{g'=1}^{r-1}{r-1\choose g'}(-1)^{g'+1}f^{g'}.
\end{equation}
For small $f$, close to the percolation transition, only the first few terms on both sides of the equality are
relevant. Retaining up to second order
\begin{equation}
1+\frac{f}{2}\approx\frac{\lambda}{r-1}\left[(r-1)-{r-1\choose 2}f\right]
\end{equation}
which produces
\begin{equation}
f\sim \frac{2(\lambda-1)}{1+(r-2)\lambda}\sim 2(\lambda-1)
\end{equation}
clearly indicating a continuous transition, in the same universality class of regular network percolation,
which diverges at the transition with exponent 1. This result is known in the literature~\cite{Newman-clusters}.

The previous results focus on hypergraphs, but their relevance to projected networks is not explicitly
clear. To clarify this, it is sufficient to explore the properties of $\phi_{ij}(o_{ij},p)$. For
this, it is useful to have in mind the asymptotic relations 
$N\frac{\partial^2}{\partial N^2}{N\choose r}\sim \frac{N^{r-1}}{(r-2)!}$ and 
${N-1\choose r-1}\sim \frac{N^{r-1}}{(r-1)!}$. 
Inserting $p=\lambda p_c$ in Eq.~(\ref{phi_oij_homo_final}), and taking the limit $N\gg o_{ij}$, the
relation 
$\phi_{ij,\text{sparse}}(o_{ij},\lambda p_c)\equiv\phi^{(s)}_{ij}(o_{ij},\lambda)=(\lambda/N)^{o_{ij}}
e^{\lambda/N}/o_{ij}!$, which is a Poisson distribution with average $\lambda/N$ (Fig.~\ref{fig-dilute}(b)). 
Therefore, as $N$ increases, the weights on the links vanish, signalling the fact that in this dilute regime, 
the hypergraph and projected networks are virtually the same, and hyperedges are non-overlapping
asymptotically. Thus, one only needs to calculate the hypergraph percolation properties to be able to
write down the projected network percolation properties. In this sparse regime, the other distributions discussed 
above have particular forms: for the hypergraph, the distribution of hyperedges visiting a node
becomes $\zeta^{(s)}_i(\ell_i,\lambda)=(\lambda/(r-1))^{\ell_i}e^{-\lambda/(r-1)}/\ell_i!$
(poisson with average $\lambda/(r-1)$), and the strength distribution on projected 
networks with $\mathcal{P}_a$ becomes 
$\xi^{(s)}_i(s_i,\lambda)=(\lambda/(r-1))^{s_i/(r-1)}e^{-\lambda/(r-1)}/[(r-1)(s_i/(r-1))!]$
(Fig.~\ref{fig-dilute}(c)).
From these results, the meaning of $\lambda$ emerges as the parameter that measures the average
node strength of the projected network. Finally, the degree distribution can be calculated
if one keeps in mind that in the sparse limit, the probability that hyperedges overlap is
minimal, and therefore, one expects that only the minimum number of hyperedges $\ell_i\to\lceil k_i/(r-1)\rceil$
contribute to the distribution. There are subtleties present in explicitly calculating
$Q_{r-1}(k_i,\lceil k_i/(r-1)\rceil)$ and $\psi^{(s)}_i(k_i,\lambda)$ when $k_i$ is not a multiple
of $r-1$ because hyperedges are forced to overlap in this case, and thus to avoid further
details, I only write the unevaluated result 
$\psi^{(s)}_i(k_i,\lambda)=Q_{r-1}(k_i,\lceil k_i/(r-1)\rceil)(\lambda p_c)^{\lceil k_i/(r-1)\rceil}
(1-\lambda p_c)^{{N-1\choose r-1}-\lceil k_i/(r-1)\rceil}$ (Fig.~\ref{fig-dilute}(d)). However, the calculations
are not prohibitive, and are derived in detail in~\cite{Lopez-pk}.

The sparse regime close to percolation is not the only possible sparse regime. 
To be concrete, note that for $p$ close to $p_c$, the average node strength
is constant, but the average overlap on projected links scales as $N^{-1}$, so the larger the 
network, the less interaction present along the links. However, one can consider a regime in which 
$\langle o_{ij}\rangle\sim \lambda/N$ is constant, and in this regime node strength increases with $N$. 
Both of these regimes are ``sparse'' in the sense that $p$ vanishes asymptotically, but each regime has
specific properties. Generally, these sparse regimes can be defined based on any sensible property,
and lead to interesting behavior.
Finally, for the dense regime ($p$ constant), the interesting effect of growth of $\langle o_{ij}\rangle$
vs. $N$ emerges, which is a unique feature of this model, and the signature that multiway interactions 
are potentially present.
\section{Conclusions}
\label{conclusions}
In conclusion, in this article I present a model of hypergraphs and associated weighted projected
networks that offers a concise and intuitive picture of hypergraphs, networks, and weights. By using 
statistical mechanics concepts, together with combinatorial tools, I have been able to determine
some basic features of homogeneous and heterogeneous projected networks that offer concrete tests
to determine whether a network that has been empirically measured may bear the signature of
multiway (group) interactions. The general idea of using the projection of a hypergraph onto a
network, which has not been studied systematically to the author's knowledge until this article, 
deserves a close look to determine further properties
that can help give a better understanding of the genuine limits and virtues of pairwise simplifications
in network research.

The author thanks 
L. Roberts,
A. Gerig, 
F. Reed-Tsochas, 
and
O. Riordan, 
for helpful discussions, and TSB/EPSRC grant SATURN (TS/H001832/1), ICT eCollective EU project (238597),
and the James Martin 21st Century Foundation Reference no: LC1213-006
for financial support. 

\clearpage
\begin{table} 
\begin{tabular}{|l||p{6cm}|p{4cm}|c|}
\hline
Set notation&Explanation&Type of element&Size\\\hline\hline
$\TboldNrij$&Hyperedges of complete hypergraph simultaneously visiting $i$ and $j$&hyperedge&${N-2\choose r-2}$ \\\hline
$O_{ij}(\sigmabold)$&Hyperedges of configuration $\sigmabold$ simultaneously visiting $i$ and $j$&hyperedge&$o_{ij}$\\\hline
$\Oboldij$&Collection of all possible sets $O_{ij}(\sigmabold)$
&Set of cardinality $o_{ij}$ of hyperedges&${{N-2\choose r-2}\choose o_{ij}}$\\\hline
\end{tabular}
\caption{Notation used for calculation of $\phi_{ij}(o_{ij},\pbold)$. The complement sets
$\overline{O}_{ij}(\sigmabold)$ are with respect to $\TboldNrij$.}
\label{table-phi_oij}
\end{table}
\begin{table} 
\begin{tabular}{|l||p{6cm}|p{4cm}|c|}
\hline
Set notation&Explanation&Type of element&Size\\\hline\hline
$\TboldNri$&Hyperedges of complete hypergraph visiting $i$&hyperedge&${N-1\choose r-1}$ \\\hline
$\lambda_i(\sigmabold)$&Hyperedges of configuration $\sigmabold$ visiting $i$&hyperedge&$\ell_i$\\\hline
$\boldsymbol{\Lambda}_i$&Collection of all possible sets $\lambda_i(\sigmabold)$
&Set of cardinality $\ell_i$ of hyperedges&${{N-1\choose r-1}\choose \ell_i}$\\\hline
\end{tabular}
\caption{Notation used for calculations of $\zeta_i(\ell_i,\pbold)$ and $\xi_i(s_i,\pbold)$ in the 
$\mathcal{P}_a$ projection. The complement sets $\overline{\lambda}_i(\sigmabold)$ are with respect to 
$\TboldNri$.}
\label{table-zeta_elli}
\end{table}
\clearpage
\begin{table} 
\begin{tabular}{|l||p{6cm}|p{4cm}|c|}
\hline
Set notation&Explanation&Type of element&Size\\\hline\hline
$K^{(\ell_i)}_i(\sigmabold)$&Hyperedges in configuration $\sigmabold$ visiting $i$ plus $k_i$ other nodes
&hyperedge&$\ell_i$\\\hline
$I^{(\rho(k_i),\ell_i)}_i(\sigmabold)$&Hyperedges in configuration $\sigmabold$ visiting $i$ plus
the $k_i$ nodes in set $\rho(k_i)$&hyperedge&$\ell_i$\\\hline
$\rho(k_i)$&Choice of $k_i$ nodes (plus $i$) in $\sigmabold$ connected to $i$ via $\ell_i$ 
hyperedges&node&$k_i$\\\hline
$\boldsymbol{\rm K}_i(\ell_i)$&Collection of all possible sets $K^{(\ell_i)}_i(\sigmabold)$
&Set of cardinality $\ell_i$ of hyperedges&${N-1\choose k_i}Q_{r-1}(k_i,\ell_i)$\\\hline
$\boldsymbol{\rm I}_i(\rho(k_i),\ell_i)$&Collection of all possible sets $I^{(\rho(k_i),\ell_i)}_i(\sigmabold)$
&Set of cadinality $\ell_i$ of hyperedges&$Q_{r-1}(k_i,\ell_i)$\\\hline
$\boldsymbol{\rm R}_i(N,k_i)$&Collection of all possible sets $\rho(k_i)$&Set of cardinality $k_i$ of nodes
&${N-1\choose k_i}$\\\hline
\end{tabular}
\caption{Notation used for calculation of $\psi_i(k_i,\pbold)$. The complement sets 
$\overline{K}^{(\ell_i)}_i(\sigmabold)$ and $\overline{I}^{(\rho(k_i),\ell_i)}_i(\sigmabold)$ are
with respect to $\TboldNri$.}
\label{table-psi_ki}
\end{table}
\clearpage
%
%

\begin{thebibliography}{99}
\bibitem{fn-bipartite} The relation between bipartite graphs and hypergraphs is well
known, and involves the duality transformation of assigning to every hyperedge an ``affiliation''
node in a bipartite graph. This relation has led to bipartite graphs also being associated in a number
of studies with projected networks of one form or another.
\bibitem{fn-other-quantities} Other quantities can be defined that would exhibit signatures of
certain rank of multiway interactions. For instance, if one defined $o_{ijg}$ as the set
of all hyperedges that simultaneously visit $i,j$ and $g$, then the same central limit
theorem argument would lead to an average that depends of ${N-3\choose r-3}$. Thus, for $r=3$
this quantity would be independent of $N$ under narrow distribution of hyperedge probabilities.
$o_{ij}$ is special because it impacts the properties of the weighted projected networks.
\end{thebibliography}
\end{document}